\documentstyle[12pt,epsfig]{article}
\begin{document}
\begin{center}
{\large{\bf Mesons and Glueballs: A Quantum Field Approach}}

\vspace*{5mm}

\underline{G.~Ganbold}\footnote{E--mail:{\tt ~~ganbold@thsun1.jinr.ru}}

\vspace*{3mm}

{\sl Bogoliubov Lab. Theor. Phys., JINR, 141980, Dubna, Russia} \\
{\sl Institute of Physics and Technology, 210651, Ulaanbaatar, Mongolia}
\end{center}

\begin{abstract}
The spectrum of two-particle bound states is investigated within a
relativistic quantum-field model of interacting quarks and gluons
confined analytically. The hadronization process of mesons and
glueballs is described by using the Bethe-Salpeter equation. Provided
by a minimal set of physical parameters (the quark masses, the
coupling constant and the confinement scale), the model satisfactorily
describes the meson ground-states, orbital and radial excitations in
a wide range from $140 \mbox{\rm MeV}$ to $9.5 \mbox{\rm GeV}$. The
estimated values for the coupling constant and the lowest-state
glueball mass are in reasonable agreement with experimental data.
\end{abstract}

\section{Introduction}

Conventionally, the observed colorless hadrons are considered as
bound states of quarks and gluons under the {\sl color confinement}
of the QCD. The quark-antiquark pair fits in a pattern, where they
have quantum numbers of the most of mesons. The color confinement
is achieved by taking into account nonperturbative and nonlinear
gluon interaction and the coupling becomes stronger in the hadron
distance \cite{pros03}. The correct summation of the higher-order
contributions becomes a problem. Besides, the structure of the QCD
vacuum is not well established yet.

The conventional QCD encounters a difficulty by defining the explicit
quark and gluon propagator at the confinement scale. Particularly,
the Schwinger-Dyson equation is used to obtain an effective quark
propagator but this implies additional assumptions on the behaviour
of the gluon propagator and quark-gluon vertices and results in
elaborated numerical evaluations \cite{mari03}.

A number of phenomenological approaches is devoted to the hadron
spectroscopy. Some of them are adapted to the specific sector of
heavy hadrons and introduce many parameters, including nonphysical
ones. The 'potential' models use a stationary nonrelativistic
Schr\"odinger equation with an increasing potential to describe the
hadronization process of few-body systems. However, the binding
energy in hadrons is not negligible, e.g., the observable two-quark
bound states have masses much heavier than the predicted combined
masses of the 'current-quarks' and it requires a relativistic
consideration.

It seems reasonable to develop a simple relativistic quantum field
model of interacting quarks and gluons and consider the formation
and spectra of the hadrons by using the Bethe-Salpeter equation.
There exists a conception of the {\sl analytic confinement} based
on the assumption that the QCD vacuum is realized by the vacuum
selfdual gluon field  which is the classical solution of the
Yang-Mills equation \cite{leut81,efned95}. Accordingly, each
isolated quark and gluon is confined and the corresponding
propagators are entire analytic functions in the background gluon
field. However, direct use of these propagators leads to long and
complicated estimations.

In our previous investigation, we have clarified the role of the
analytic confinement in formation of the hadron bound states by using
simple scalar-field models \cite{efim02}. We have considered "mesons"
composed from two "scalar quarks" glued by "scalar gluons" by omitting
the spin, color and flavor degrees of freedom. Despite the simplicity,
these models resulted in a quite reasonable sight to the underlying
physical principles of the hadronization mechanism and could give a
qualitative description of the "scalar" mesons \cite{ganb04}.

Below we extend the consideration by taking into account the spin,
color and flavor degrees of freedom for constituent quarks and gluons
within a simple relativistic quantum field model. This allows us to
obtain extended and accurate results not only for the meson ground
states but also for the orbital and radial excitations as well as the
coupling constant and the glueball lowest state.

\section{The Model}

Consider a Yukawa-type interaction of the quarks $\Psi_{\alpha f}^a(x)$
and gluons $\phi^a_\mu(x)$ confined analytically. The model Lagrangian reads
\cite{ganb05}
\begin{eqnarray}
{\cal L}\!=\! \left({\bar\Psi_\alpha^i}[S^{-1}]_{\alpha\beta}^{ij}
{\Psi_\beta^j}\right)
\!+\! {1\over 2}\left(\phi_\mu^a[D^{-1}]_{\mu\nu}^{ab} \phi_\nu^b \right)
\!+\! g\left({\bar\Psi_\alpha^i} [i\gamma_\mu]^{\alpha\beta}
      t^a_{ij}{\Psi_\beta^j}\phi_\mu^a \right)
\!+\! g\left(\phi^a_\mu \phi^b_\nu F^c_{\mu\nu} \right) f^{abc},
\end{eqnarray}
where $g$ - the coupling constant,
$F^c_{\mu\nu}=\partial_\mu \phi_\nu^c-\partial_\nu \phi_\mu^c$ and
$\{a,\alpha,f\}$ - the color, spin and flavor indices.

We use a minimal set of physical parameters, namely, the coupling
constant, the scale of confinement and the quark masses:
$\{\alpha_s=g^2/4\pi,\Lambda,m_{ud},m_s,m_c,m_b\}$. Hereby, we do not
distinct the masses of $u$ and $s$ quarks and neglect the superheavy
$t$ quark due to the absence of observed ($t\bar{t}$) bound states.

The matrix elements of hadron processes at large distance are in
fact integrated characteristics of the quark and gluon propagators
and the interaction vertices. Taking into account the correct global
symmetry properties and their breaking by introducing additional
physical parameters may be more important than the working out in
detail (e.g., \cite{feld00}). We admit that the bound states of quarks
and gluons may be found as the solution of the Bethe-Salpeter equation
in a variational-integral form \cite{ganb05}. One may guess that
the solution is not too sensible on the details of integrands.

We consider the effective quark and gluon propagators as follows:
\begin{eqnarray}
&& \tilde{S}_{\alpha\beta}^{ij}(\hat{p})=
~\delta^{ij} {\left\{ i\hat{p}+m~[1+\gamma_5\omega(m)]
\right\}_{\alpha\beta}\over m^2}
\exp\left\{-{p^2+m^2\over 2\Lambda^2} \right\}\,, \nonumber\\
&& D^{ab}_{\mu\nu}(x)=\delta^{ab}\delta_{\mu\nu}~{\Lambda^2\over (4\pi)^2}
~\exp\left\{-{x^2\Lambda^2\over 4}\right\}
= \delta^{ab}\delta_{\mu\nu}~D(x)  \,,
\label{propagat}
\end{eqnarray}
where $\hat{p}=p_\mu \gamma_\mu$, $m$ - the quark mass and
$0<\omega(m)<1$. These entire analytic functions in Euclidean
metric represent more realistic and reasonable extensions to model
propagators used in our earlier investigations
\cite{efim02,ganb04,ganb05} and allow us to estimate the meson and
glueball spectra with sufficient accuracy by avoiding cumbersome
calculations.

Consider the partition function
\begin{eqnarray}
Z=\int\!\!\!\int\!\!\delta\bar\Psi\delta\Psi \int\!\!\delta\phi
\exp\left\{-(\bar\Psi S^{-1}\Psi)-{1\over 2}(\phi D^{-1}\phi)
-g(\bar\Psi\Gamma\Psi\phi)-g(\phi\phi F)\right\} \,.
\end{eqnarray}

We suppose, that the coupling $\alpha_s$ is relatively weak (in
Section 3.1 we show that the coupling is indeed small). Then, we can
restrict the consideration within the one-gluon exchange approximation
and take the Gaussian path integration over  $\phi$-variable:
\begin{eqnarray}
\int\!\!\delta\phi
\exp\left\{-{1\over 2}(\phi D^{-1}\phi)-g(\bar\Psi\Gamma\Psi\phi)
-g(\phi\phi F)\right\}
=\exp\left\{-{\cal L}_2 [\bar\Psi,\Psi]-{\cal L}_G\right\} \,,
\end{eqnarray}
where, the terms corresponding to the two-quark and two-gluon bound states
read
\begin{eqnarray}
{\cal L}_2 ={g^2\over 2} \int\!\!\delta\phi ~e^{-{\cal L}_B[\phi]}
\left( (\bar\Psi\Gamma\Psi) D (\bar\Psi\Gamma\Psi) \right) \,,
\qquad
{\cal L}_G ={27 g^2\over 2} \int\!\!\delta\phi ~e^{-{\cal L}_B[\phi]}
\left( \phi \phi D \phi \phi \right)\,.
\label{gluestate}
\end{eqnarray}

\section{Mesons}

Describe shortly the important steps of our approach on the example
of the two-quark bound states, more details can be found in
\cite{efim02,ganb04}. First, we allocate the one-gluon exchange between
quark currents, go to the relative co-ordinates in the "center-of-mass"
system and introduce the relative mass of quark $\mu_f=m_f/(m_1+m_2)$.
The latter is important for mesons consisting of different quark flavors.
The next step is to perform a Fierz transformation to
get the colorless bilocal quark currents. Then, introduce an normal
basis $\{U(x)\}$. The appropriate diagonalization of ${\cal L}_2$
on the colorless quark currents should be done by using $\{U(x)\}$.
Use a Gaussian path-integral representation by introducing auxiliary
meson fields $B_{\cal N}$ and apply the 'Hadronization Ansatz' to
identify $B_{\cal N}(x)$ with meson fields carrying the quantum numbers
${\cal N}=\{Jff'\}$, where $J$ - the spin and $f$ - the quark flavor.

Below we consider the pseudoscalar  ($P:~J^{PC}=0^{-+}$) and vector
($V:~J^{PC}=1^{--}$) mesons, the most established sectors of hadron
spectroscopy.

The partition function for mesons reads:
\begin{eqnarray}
Z_{\cal N}=\int\!\! DB_{\cal N}~\exp\left\{ -{1\over 2}(B_{{\cal N}}
~[1+g^2{\bf Tr}(V_{\cal N}~S~V_{\cal N}~S)]~B_{{\cal N}})
+W_I[B_{{\cal N}}]\right\}\,,
\end{eqnarray}
where $V_{{\cal N}}=\Gamma_J\int dy~U(y)\sqrt{D(y)} \exp\{{y \mu_f \stackrel
{\leftrightarrow}{\partial}}\}$ is a vertice function, $J=\{S,P,V,A,T\}$
and
$\Gamma_J=\{I,i\gamma_5,i\gamma_\mu,i\gamma_5\gamma_\mu,\sigma_{\mu\nu}\}$.
We use a Euclidean metric, with:
$\{\gamma_\mu, \gamma_\nu\} = 2\delta_{\mu\nu}$;
$\gamma_\mu^{\dag} = \gamma_\mu$; $ab =\sum_{i=1}^{4}{a_i b_i}$.
For a timelike vector $p_\mu$, $p^2 < 0$.

The final-state interaction between mesons is described by
$W_I[B_{\cal N}]=O[B_{\cal N}^3]$.

The diagonalization of the quadratic form on the orthonormal system
$\{U_{\cal N}\}$ is nothing else but the solution of the
Bethe-Salpeter equation (in the one-gluon approximation):
\begin{eqnarray}
&& 1+g^2{\bf Tr}(V_{{\cal N}}SV_{{\cal N'}}S)
= \int\!\!\!\!\int\! dx dy
~U(x)\left\{ 1+g^2\sqrt{D(x)}\int\!
{d^4 k\over(2\pi)^4}~e^{-ik(x-y)}  \right. \nonumber\\
&& \left. \cdot
{\bf Tr}\left[\Gamma_J
\tilde{S}\left(\hat{k}+\mu_1\hat{p}\right)\Gamma_{J'}
\tilde{S}\left(\hat{k}-\mu_2\hat{p}\right)
\right]\sqrt{D(y)} \right\}
U(y)=\delta^{{\cal NN'}} [1+\lambda_{{\cal N}}(-p^2)] \,.
\label{Bethe1}
\end{eqnarray}

Then, the meson mass is derived from equation:
\begin{eqnarray}
1+\lambda_{{\cal N}}(M_{{\cal N}}^2)=0\,.
\label{Bethe2}
\end{eqnarray}

\subsection{Ground States}

We should solve the Bethe-Salpeter equation (\ref{Bethe1}) with
sufficient accuracy for the meson ground states. The polarization
kernel is real and symmetric that allows us to find a simple
variational solution to this problem.

According to (\ref{Bethe1}) it is reasonable to choose a normalized
trial function \cite{efim02,ganb04,ganb05}:
\begin{eqnarray}
U_{\ell\{\mu\}}(x,a)\sim T_{\ell\{\mu\}}(x)\sqrt{D(x)}~
e^{-a{\Lambda^2 x^2\over 4}}\,, \qquad
\sum\limits_{\{\mu\}}\int dx \left| U_{\ell\{\mu\}}(x,a)\right|^2 =1\,,
\label{trial1}
\end{eqnarray}
where $\ell$ is the orbital quantum number, $T_{\ell\{\mu\}}(x)$
is the four-dimensional spherical harmonic and $a>0$ is a parameter.

Substituting (\ref{trial1}) into (\ref{Bethe1}) we obtain a variational
equation for the meson mass
\begin{eqnarray}
0 = 1+\max\limits_{a>0}\sum\limits_{\{\mu\}}
\int\!\!\!\!\int\! dx dy~
U_{\ell\{\mu\}}(x,a)
~g^2\Pi_p(x,y)
~U_{\ell\{\mu\}}(y,a)\,, \quad p^2=-M_\ell^2 \,.
\label{variat}
\end{eqnarray}

\begin{figure}[ht]
 \centerline{
 \includegraphics[width=70mm,height=60mm]{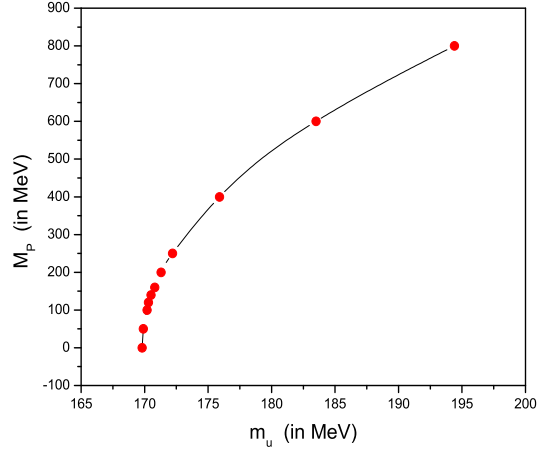} }
 \caption{Evolution of the pseudoscalar meson mass with
 quark mass $m_u$. }
\end{figure}

Below we consider $\ell=0$, the pseudoscalar ($J^{PC}=0^{-+}$) and
vector ($1^{--}$) mesons, the most established sectors of hadron spectroscopy.

For the ground state mass $M=M_0$ we obtain:
\begin{eqnarray}
1 &=&  {\alpha_s \Lambda^4\over 48\pi m_1^2 m_2^2}
~\exp\left\{{M^2 (\mu_1^2+\mu_2^2) -m_1^2-m_2^2 \over 2\Lambda^2} \right\}
\max\limits_{1/3< b <1} \left\{ \left[ {(5\,b-1)(1-b)\over b}\right]^2
\right. \nonumber\\
&& \!\!\!\!\! \cdot \exp\left[-{b\,M^2(\mu_1-\mu_2)^2\over 4\Lambda^2}\right]
\left[ 2\,b\,\rho_J +{M^2\over \Lambda^2} \left(\mu_1\,\mu_2 + b\,(2-b\,\rho_j)
{(\mu_1-\mu_2)^2\over 4}\right) \right. \nonumber\\
&& \!\!\!\!\! \left. \left. + {m_1 m_2\over \Lambda^2}
\left( 1+\chi_J ~\omega\left({m_1\over\Lambda}\right)
\omega\left({m_2\over\Lambda} \right) \right) \right] \right\} \,,
\label{ground}
\end{eqnarray}
where $\omega(z)=1/(1+z^2/4)$ and
$$
\rho_J=\left\{1,1/2\right\}\,,~~~
\chi_J=\left\{+1,-1\right\} ~~~ \mbox{\rm for}~~~J=\{P,V\}\,.
$$

Equation (\ref{ground}) is nontrivial, the behaviour of the quark
propagator $\tilde{S}\sim 1/m^2$ imposes a restriction on the value
of the quark mass from below. Particularly, the evolution of the
pseudoscalar meson mass $M_P(m_u,m_u)$ with "constituent-quark"
mass $m_u$ at fixed values of $\{\alpha_s,\Lambda\}$ is depicted
in Fig.1. The optimal value of $m_u$  is fixed from the relation
$M_P(m_u,m_u)=M_\pi=138$ MeV.

By comparing our estimates with the observed meson masses
\cite{PDG2006} we have found the following optimal values of free
parameters \cite{ganb05}:
\begin{eqnarray}
\alpha_s=0.186 \,,\qquad  \Lambda=730 {\mbox{\rm ~MeV}} \,,
\qquad m_{ud}=170{\mbox{\rm ~MeV}} \,, \nonumber\\
m_s=188{\mbox{\rm ~MeV}}  \,,\qquad m_c=646{\mbox{\rm ~MeV}}  \,,
\qquad m_b=4221{\mbox{\rm ~MeV}} \,.
\label{parameters}
\end{eqnarray}

\begin{table}[h]
\begin{tabular}{|c|c|c|c|c|c|c|c|}
 \hline
               &     &               &     &                &     &               &     \\
$J^{PC}=0^{-+}$&$M_P$&$J^{PC}=0^{-+}$&$M_P$&$J^{PC}=1^{--}$ &$M_V$&$J^{PC}=1^{--}$&$M_V$\\
               &     &               &     &                &     &               &     \\
 \hline
 $\pi(138)$    & 138 & $D(1870)$     & 1928& $\rho(770)$    & 813 &$K^*(892)$     & 927 \\
 $\eta(547)$   & 508 & $D_s(1970)$   & 2009& $\omega(782)$  & 826 &$D^*(2010)$    & 2028\\
 $\eta_c(2979)$& 3018& $B(5279)$     & 5359& $\Phi(1019)$   & 1041&$D^*_s(2112)$  & 2106\\
 $\eta_b(9300)$& 9458& $B_s(5370)$   & 5397& $J/\Psi(3097)$ & 3097&$B^*(5325)$    & 5361\\
 $K(495)$      & 495 & $B_c(6286)$   & 6074&$\Upsilon(9460)$& 9460&               &     \\
\hline
\end{tabular}
\caption{Estimated ground-state masses $M_P$ and $M_V$ (in {\mbox{\rm MeV}}).}
\end{table}

Our estimates for the pseudoscalar and vector meson masses in the
ground state (Tab.1 and Fig.2) compared with experimental data show
that the relative error is small ($\sim 1\div 5$ per cent)
in a wide range of mass $\sim 140\div 9500{\mbox{\rm ~MeV}}$.

\begin{figure}[ht]
 \centerline{
 \includegraphics[width=100mm,height=80mm]{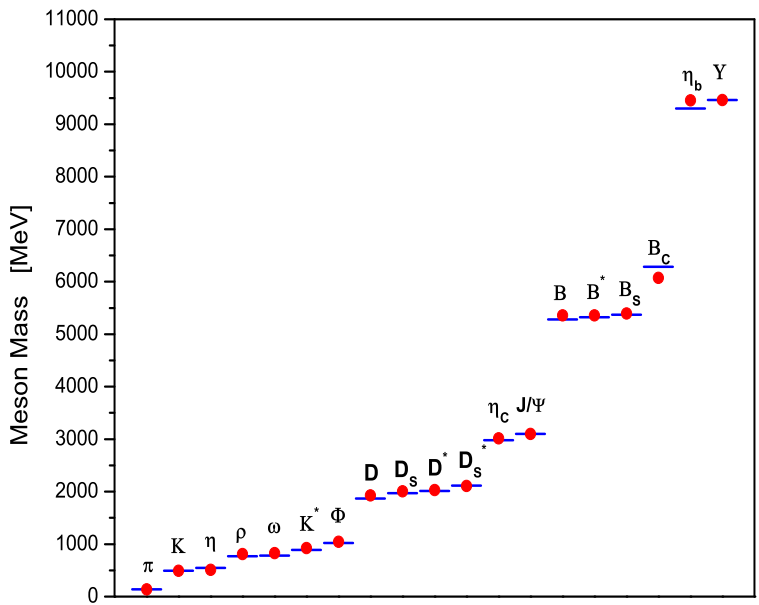} }
 \caption{The estimated meson masses (dots) and the experimental data (dashes).}
\end{figure}

We note:
\begin{enumerate}
\item
The set of optimal parameters (\ref{parameters}) is not unique,
another set also gives a reasonable ground-state spectra for $P$
and $V$ mesons. But we choose these values by taking into account
other important hadron data such as the orbital and radial excitations
of mesons as well as the lowest-state glueball mass.
\item
The obtained value $\alpha_s=0.186$ is in agreement with the latest
experimental data for the QCD coupling value
$\alpha_s^{QCD}\approx 0.10\div 0.35$
on the hadron distance \cite{PDG2006} and with the prediction of the
quenched theory $\alpha_s^{quench}\approx 0.195$ \cite{kasz05}. Moreover,
this relatively weak coupling value justifies the use of the one-gluon
exchange mode in our consideration. A possible use of a "sliding"
interaction strength $\alpha_s(M)$ depending on the mass scale
(see, e.g., \cite{solo02,nest03,kasz05}) will be discussed later.
\item
The obtained quark masses fit an hierarchy: $m_{u,d} < m_s < m_c < m_b$.
\item
Our model provides the $SU(3)$-symmetry breaking: $M_{\pi} \not= M_K$.
\item
The $\omega$ and $\Phi$ mesons considered as mixed states of
$u\bar{u}+d\bar{d}$ and $s\bar{s}$ pairs with {\sl a priori} unknown
mixing angle $\theta_V$. The optimal value is found: $\theta_V=9^\circ$.
\item
The $U(1)$-splitting is explained, i.e., we obtain a large mass
difference $M_\pi/M_\eta\approx 0.25$ between pseudoscalar isovector
and isoscalar mesons, while $M_\rho / M_\omega \approx 1$ for the
vector isovector and isoscalar.
\item
The dependence of the meson masses $M(m_1+m_2)$ on the sum of two
"constituent" quark masses (Fig.3) is nontrivial.
\item
A rough estimate of the hadronization region
$$
r_{hadr}^2 \sim {\int d^4 x~x^2~D(x) \over\int d^4
x~x~D(x)}={8\over\Lambda^2} \sim \left({1\over 250 \mbox{\rm MeV}}\right)^2
$$
shows that the confinement scale $r_{conf}\simeq 1/\Lambda$ is comparable
with $r_{hadr}$.
\end{enumerate}

\begin{figure}[ht]
 \centerline{
 \includegraphics[width=70mm,height=60mm]{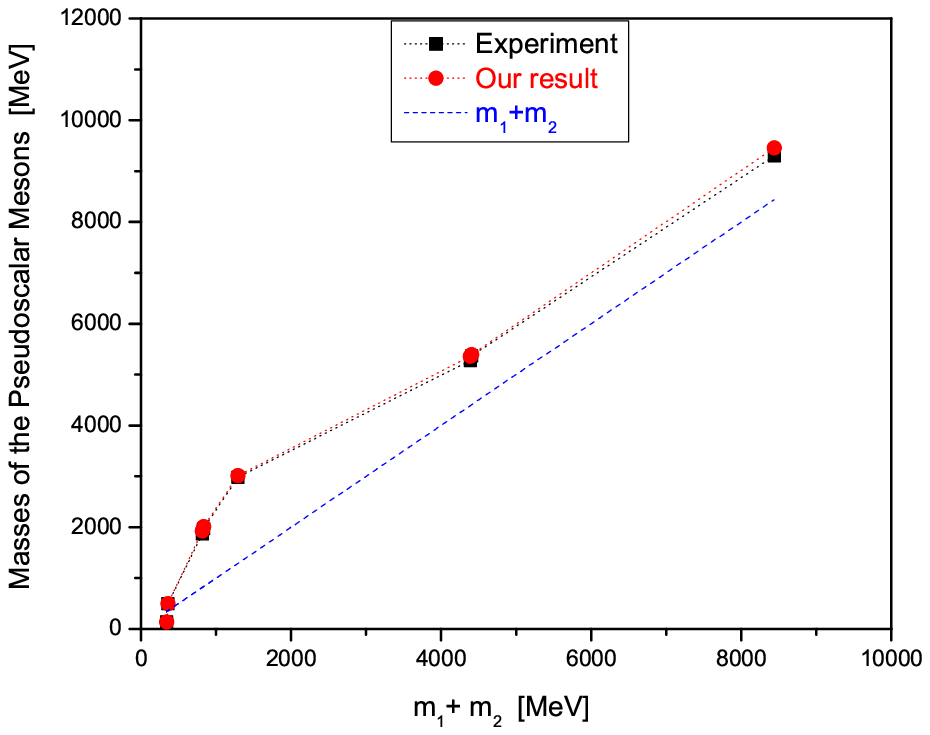}
 \includegraphics[width=70mm,height=60mm]{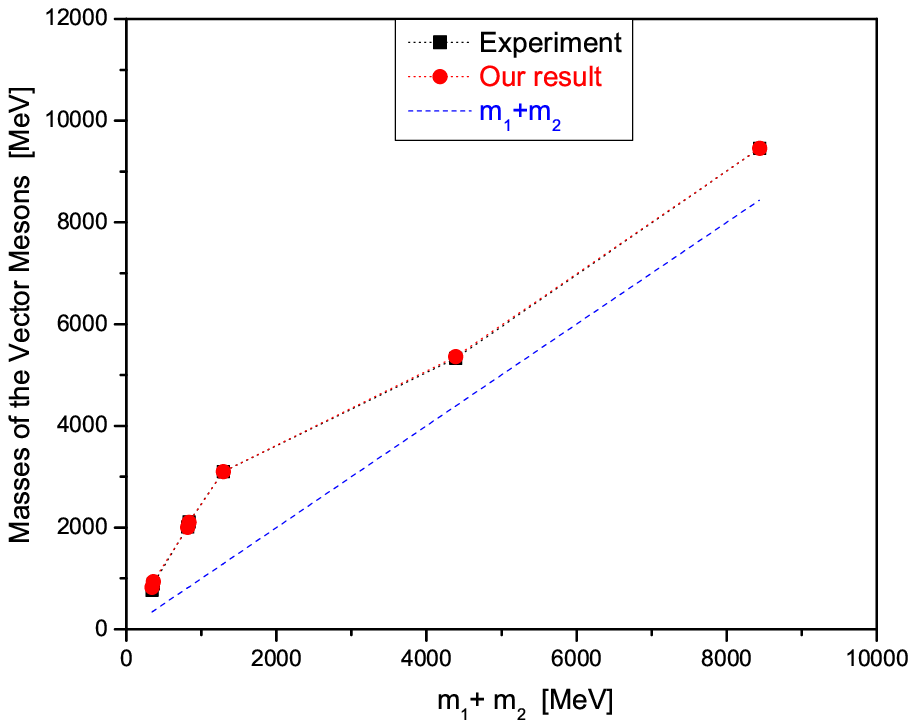} }
 \caption{The estimated masses for the $P$ and $V$ mesons versa
 the sum of quark masses. The dashed straight lines depict the plain
 sum $m_1+m_2$.}
\end{figure}

Below, we extend our consideration to the orbital ($\ell > 0$) and
radial ($n_r > 0$) excitations of mesons.

\section{Meson Regge Trajectories with $n_r=0$}

The orbital excitations take place in larger distances and should
be less sensitive to the short-range details of the propagators.
Therefore, correct description of the meson Regge trajectories can
serve as an additional test ground for our model.

Below we concentrate mostly on the orbital excitations with zero
radial number $n_r=0$ for which experimental data exist for all
states with $\ell\le 5$.

Particularly, by substituting (\ref{trial1}) into (\ref{Bethe1})
we obtain the mass equation:
\begin{eqnarray}
1 \!\!\!
&=& \!\!\! {\alpha_s \Lambda^4\over 48\pi m_1^2 m_2^2}
{1\over 4^{\ell}({\ell}+1)!}~\exp\left\{{(m_1^2+m_2^2)\over 2\Lambda^2}
\left[{M_{\ell}^2\over (m_1+m_2)^2}-1 \right]\right\}            \nonumber\\
&& \!\!\!\!\! \cdot \max\limits_{1/3<c<1} \left\{
\left[ {(5\,c-1)(1-c)\over c^2}\right]^2~ \left(c^2 {d\over dc} \right)^{\ell}
\left\{ c^2 \exp\left[-{c\,M_{\ell}^2(m_1-m_2)^2\over 4\Lambda^2 (m_1+m_2)^2}
\right] \right. \right.\nonumber\\
&&  \!\!\!\!\!
\cdot \left[ 2\,c\,\rho_J +{M_{\ell}^2\over \Lambda^2 (m_1+m_2)^2}
\left(m_1\, m_2 + {c\,(2-c\,\rho_j)\over 4}~(m_1 - m_2)^2\right) \right. \nonumber\\
&&  \!\!\!\!\! \left. \left. \left. +{m_1\,m_2\over \Lambda^2}
\left( 1+\chi_J ~\omega\left({m_1\over\Lambda}\right)
\omega\left({m_2\over\Lambda} \right) \right) \right] \right\}  \right\}\,,
\label{orbital}
\end{eqnarray}
where $\omega(z)=1/(1+z^2/4)$ and
$ \rho_J=\left\{1,1/2\right\}\,,~~
\chi_J=\left\{+1,-1\right\} ~~ \mbox{\rm for}~~J=\{P,V\}\,$.

\begin{figure}[ht]
 \centerline{
 \includegraphics[width=70mm,height=60mm]{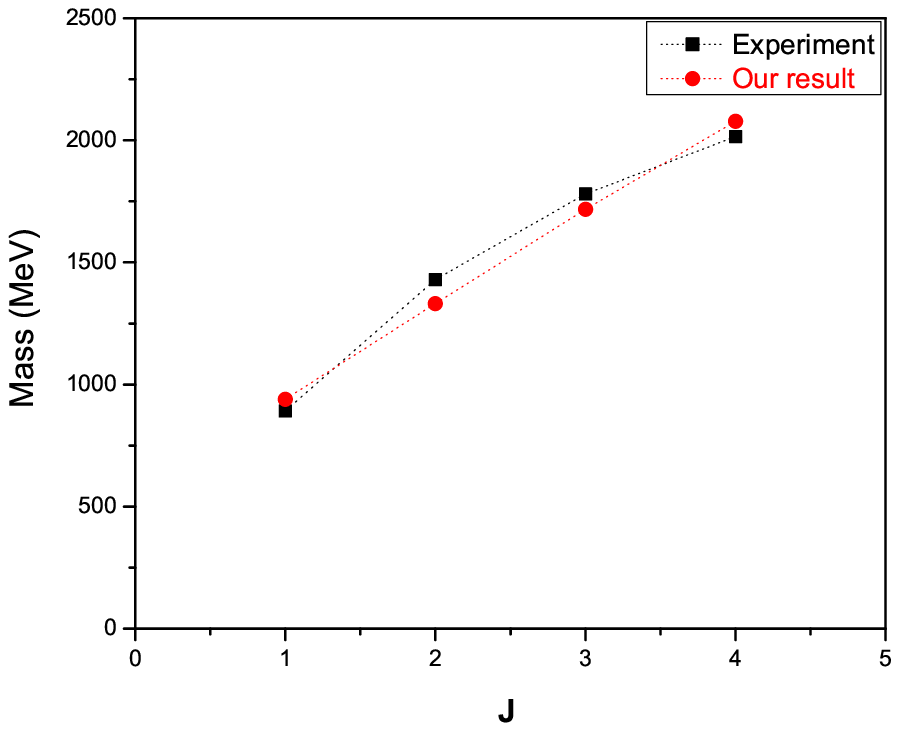}
 \includegraphics[width=70mm,height=60mm]{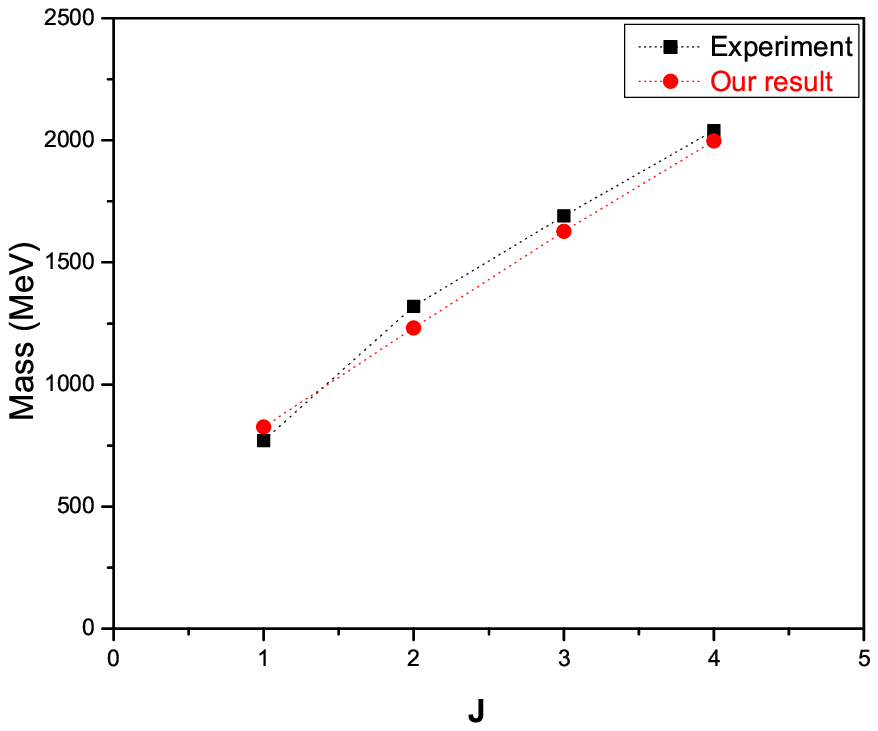} }
 \caption{The estimated mass $M_J$ (dots) for the $\rho$- and
 $K^*$-meson excitations.}
\end{figure}

Substituting the optimal parameters (\ref{parameters}) into
(\ref{orbital}) we have estimated the masses of excitations for the
$\rho$-meson and $K$-meson families. Our estimates (dots) plotted
versa the quantum number $J=\ell+s$ ~($s=\{0,1\}$ - the spin of
the $(q\bar{q'})$ pair) is given in Fig.4 in comparison with the
experimental data (boxes) \cite{PDG2006}.

As is expected, our model describes acceptably the $J$-trajectories
of the $\rho$-meson and $K$-meson excitations.

For large $\ell\gg 1$ the asymptotical behaviour of $M_\ell^2$ reads
\begin{eqnarray}
M_{\ell}^2 \simeq 4(m_1+m_2)^2+4\Lambda^2(\ell+2)
\ln\left( {2\over 3-\sqrt{5}}\right) = M_{0}^2 + \ell\cdot\Omega_{\ell} \,.
\label{reggeL1}
\end{eqnarray}

We note that for large $J\gg 1$:

i) The Regge trajectories of $K^*$ and $\rho$ mesons are linear and similar.

ii) The intercept $M_{0}^2$ is positive and depends on quark masses and $\Lambda$.

ii) The slope $\Omega_{\ell}$ sensitively depends only on $\Lambda$.

\vskip 3mm

\section{Meson Radial Excitations $n_r>0$}

A nice example of the meson radial excitations is the "charmonium" sector.
Experimentally well established is the $J/\Psi$-family $(c\bar{c})$ with
 $J^{PC}=1^{--}$.

We expand the radial part of the normalized trial function as follows:
\begin{eqnarray*}
\label{testf2}
U_{n}(x)\sim L_{n}^{(1/2)}(2 b \vec{x}^2)
\exp\left\{-\Lambda^2(b \vec{x}^2+c x_0^2)\right\} \,, \qquad
\int\! dx \left| U_{n}(x)\right|^2 =1 \,, \qquad x=\{x_0,\vec{x}\} \,,
\end{eqnarray*}
where $\{b,c\}$ -- dimensionless positive parameters, $n$ -- the radial
quantum number and $L_n^{(1/2)}(z)$ is the generalized Laguerre polynomial.

Taking into account the orthogonality of $L_n^{(1/2)}(z)$ on the interval
$(0,\infty)$ with respect to the weight function $w(z)=z^{1/2}~e^{-z}$
we arrive in the equation for the radial-excitation spectrum:
\begin{eqnarray}
\label{radial}
1 \!\!\! &=& \!\!\! {\alpha_s \over 48 \sqrt{\pi}} \left({\Lambda\over m}\right)^4
\exp\left\{ {M_n^2 \over 4\Lambda^2}-{m^2 \over \Lambda^2} \right\}
~{1\over n!(n+1)!} ~\max\limits_{A,B} \left\{ \left[{ {(5A-1)(1-A)\over A}}
\right]^{1/2}  \right. \nonumber \\
&& \!\!\!\!\! \left.  \cdot ~(2B-1)^{3/2}
{d^{2n}\over dt^n ds^n} \left[{A+{3\,u\,v\over u\,v+u+v}+{M_n^2\over \Lambda^2}
+{4\,m^2\over\Lambda^2}(1-\omega^2({m/\Lambda}))\over [(1-t)(1-s)(u\,v+u+v)]^{3/2}}
\right] \right\}_{t=0,~s=0} \,,
\end{eqnarray}
where
$$
u=B+{(2B-1)\,s\over (1-s)}\,,~~~~v=B+{(2B-1)\,t\over (1-t)}\,,~~~~~~
{1\over 3}<A<1<B<\infty \,.
$$

The numerical solutions of Eq.(\ref{radial}) for $M_n(c\bar{c})$ are
plotted in Fig.5. We see that there exists an expressed convexity for
$n<3$ and for higher mass members the $n$-trajectory is approximately
linear. By analysing (\ref{radial}) for large $n$ we obtain the solution:
\begin{eqnarray}
M_{n}^2(m,\Lambda) = M_{0}^2(m,\Lambda) + n\cdot\Omega_n(\Lambda) \,.
\label{reggeN1}
\end{eqnarray}

For $n\gg 1$:

i) The radial Regge trajectories are linear.

ii) The intercept $M_{0}^2$ depends on both $m$ and $\Lambda$.

ii) The slope $\Omega_n$ depends only on $\Lambda$.

\begin{figure}[ht]
 \centerline{
 \includegraphics[width=70mm,height=60mm]{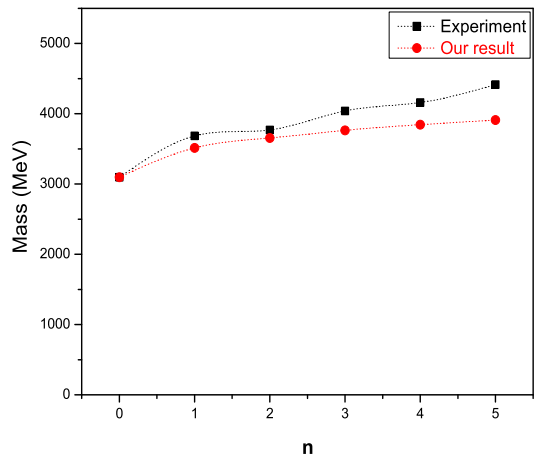} }
 \caption{The spectrum of the $J/\Psi$-family radial excitations
  for $\Lambda=730~{\mbox{\rm MeV}}$.}
\end{figure}

Note, for the $1^{--}$ radial excitations of charmonium, we obtain
a correct (convex, asymptotically linear) behaviour, but underestimate
the experimental data. For larger $\Lambda\sim 800\div850~{\mbox{\rm MeV}}$
our estimate approaches the results in \cite{PDG2006}.

\section{Glueball Lowest State}

In contrast to the meson sector, the experimental status of the
glueball is not clear. The signatures naively expected for glueballs
in the experiment are: the absence in any ($q\bar{q}$) nonets,
an enhanced production in gluon-rich channels of central productions
and radiative decays, a decay branching fraction incompatible with
two-quark states and the reduced couplings.

There are predictions expecting non-$q\bar{q}$ scalar objects, like
glueballs and multiquark states in the mass range
$\sim 1600\div 1800{\mbox{\rm ~MeV}}$ \cite{amsl04,bugg04}. Recent
lattice calculations, QCD sum rules, "tube" and constituent glue
models agree that the lightest glueballs have quantum numbers
$J^{PC}=0^{++}$ and $2^{++}$.

A QCD sum rule \cite{fork05} and the K-matrix analysis \cite{anis04}
predict the scalar glueball mass $M_G=1.25\pm 0.2{\mbox{\rm ~GeV}}$
reducing down the quantized knot soliton model result \cite{kond06}
and the earlier quenched lattice estimates $M_G\simeq 1550{\mbox{\rm ~MeV}}$
\cite{bali93,morn99,wein00}. However, most recent quenched lattice
estimate with improved action favors a larger mass close to
$M_G=1710\pm 50 \pm 58 {\mbox{\rm ~MeV}}$ \cite{chen06}.

It is evident that the structure of QCD vacuum plays the main role
in the formation of glueballs. We suppose that the lowest state
of the glueball may be reasonably described by the "gluon-gluon"
bound state $J^{PC}=0^{++}$ in our model.

The Gaussian character of the gluon propagator (\ref{propagat})
allows us explicitly solve the equation for the glueball mass
and the result reads:
\begin{eqnarray}
\label{glueball}
M_G^2 = 2\Lambda^2\ln\left({\alpha_{upp}\over\alpha_s}\right) \,,
\qquad \alpha_{upp}={2\pi(2+\sqrt{3})^2\over 27} \,.
\end{eqnarray}

Note, the glueball mass (\ref{glueball}) depends on $\alpha_s$ in a
nonperturbative way and the physical restriction $M_G^2 \ge 0$ leads
to an upper bound: $\alpha_s < \alpha_{upp}\approx 3.24124$.

Substituting the optimal parameters
($\alpha_s=0.186,~\Lambda=730{\mbox{\rm ~MeV}} $) obtained for the meson
sector into (\ref{glueball}) we estimate the lowest-state scalar glueball
mass
\begin{eqnarray}
M_G = 1745{\mbox{\rm ~MeV}} \,.
\label{Glueballmass}
\end{eqnarray}

Our estimate (\ref{Glueballmass}) does not contradict the latest
predictions expecting glueballs in the mass range
$\sim 1600\div 1800{\mbox{\rm ~MeV}}$ \cite{amsl04,bugg04}.

\vskip 5mm

In conclusion, we have considered a relativistic quantum field model
of interacting quarks and gluons under the analytic confinement. We
have used only physical parameters (the quark masses, the coupling
constant and the confinement scale) for the model and solved the
Bethe-Salpeter equation in the ladder approximation for the hadron
bound states. The use of one-gluon exchange mode is justified by the
estimated small value $\alpha_s=0.186$ of the coupling constant
that is also in agreement with both the prediction of the quenched
theory $\alpha_{quenched}\approx 0.195$ \cite{kasz05} and the latest
experimental data for the QCD running coupling $\alpha_s\approx 0.10\div 0.35$.

Our approach does not require the "flux tube"-type confinement for the
light and heavy mesons.

Within a simple relativistic model with reasonable forms of the quark
and gluon propagators we describe correctly:

- the pseudoscalar and vector meson masses in the ground state and
orbital excitations. The relative error is small in a wide range
of mass $140{\mbox{\rm ~MeV}} \div 9.46{\mbox{\rm ~GeV}}$,

- the quark mass hierarchy: $m_{u,d} < m_s < m_c < m_b$,

- the SU(3)-symmetry breaking: $M_K \not= M_{\pi}$,

- the so-called "$U(1)$-splitting": $M_\pi\ll M_\eta$ while
   $M_\rho \approx M_\omega$,

- the approximately linear radial and orbital Regge trajectories
  for the pseudoscalar and vector mesons,

- the lowest state glueball mass close to $\sim 1750{\mbox{\rm ~MeV}}$.

\vskip 3mm

The author thanks V.V.Burov, G.V.Efimov, S.B.Gerasimov, E.Klempt and
W.Oelert for useful discussions and comments.



\begin{thebibliography}{99}
\bibitem{pros03} M.~Baldicchi and G.M.~Prospri, arXiv:hep-ph/0310213 (2003).
\bibitem{mari03} P.~Maris and C.D.~Roberts, Int. J. Mod. Phys., {\bf E12} (2003) 297.
\bibitem{leut81} H.~Leutwyler, Phys. Lett., {\bf 96B} (1980) 154;
                 Nucl. Phys., {\bf B179} (1981) 129.
\bibitem{efned95} G.V.~Efimov and S.N.~Nedelko, Phys. Rev., {\bf D51} (1995) 174; \\
                 Ja.V.~Burdanov et al., Phys. Rev., {\bf D54} (1996) 4483.
\bibitem{efim02} G.V.~Efimov and G.~Ganbold, Phys. Rev., {\bf D65}  (2002) 054012.
\bibitem{ganb04} G.~Ganbold, AIP Conf. Proc., {\bf 717} (2004) 285.
\bibitem{ganb05} G.~Ganbold, AIP Conf. Proc., {\bf 796} (2005) 127; arXiv:hep-ph/0512287 (2005).
\bibitem{PDG2006} Particle Data Group, W.M-~Yao et al., J. Phys., {\bf G33}  (2006) 1.
\bibitem{feld00} T.~Feldman, Int. J. Mod. Phys., {\bf A15}  (2000) 159.
\bibitem{kasz05} O.~Kaszmarek and F.~Zantow, Phys. Rev., {\bf D71} (2005) 114510.
\bibitem{solo02} D.V.~Shirkov, Theor. Math. Phys., {\bf 132} (2002) 484.
\bibitem{nest03} A.V.~Nesterenko, Int. J. M. Phys., {\bf A18} (2003) 5475.
\bibitem{amsl04} C.~Amsler, N.A.~Tornqvist, Phys. Rep., {\bf 389} (2004) 61.
\bibitem{bugg04} D.V.~Bugg, Phys. Lett., {\bf C397} (2004) 257.
\bibitem{fork05} H.~Forkel, Phys. Rev., {\bf D71}  (2005) 054008.
\bibitem{anis04} V.V.~Anisovich, AIP Conf. Proc., {\bf 717} (2004) 441; arXiv:hep-ph/0310165 (2004).
\bibitem{kond06} K.-I.~Kondo et al., arXiv:hep-th/0604006 (2006).
\bibitem{bali93} G.~Bali et. al., Phys. Lett., {\bf B309} (1993) 378;
\bibitem{morn99} C.~Morningstar and M.~Peardon, Phys. Rev., {\bf D60}  (1999) 034509.
\bibitem{wein00} W.~Lee and D.~Weingarten, Phys. Rev., {\bf D61}  (2000) 014015.
\bibitem{chen06} Y.~Chen et. al., Phys. Rev., {\bf D73} (2006) 014516.
\end{thebibliography}
\end{document}